# IMPROVED ASTABLE MULTIVIBRATOR

RAJU BADDI

National Center for Radio Astrophysics, TIFR, Ganeshkhind P.O Bag 3, Pune University Campus, PUNE 411007, Maharashtra, INDIA; baddi@ncra.tifr.res.in

**ABSTRACT**

The conventional two transistor astable multivibrator has been modified to produce better waveform which are more steeply rising and falling than those from the conventional astable. The improvement in the waveform is achieved by using a diode pair in each collector branch of the transistors. This restricts the direction of current flow for the charging and discharging of the timing capacitors. This article also derives the time period of oscillations for the modified circuit and discusses the limits on source and sink currents at the transistor collectors.

## I. INTRODUCTION

The conventional two transistor astable(Giacoletto 1977; Kasatkin & Nemtsov 1986) as shown in Figure 1 produces alternating voltage levels at the collectors of the transistors marked T1 and T2. This change of voltage levels is characterized by an approximate rectangular wave and is shown to the right of the circuit diagram. The conventional two transistor astable suffers from slow rising(npn version) and slow falling(pnp version) output waveforms. This happens due to the charging (blue line) of the capacitors which connect to the collector of the transistors. Due to finite time required by the capacitor to charge to the full voltage allowed by the branch the voltage at the collector does not increase sharply and exhibits a slow rise in the output waveform(Figure 1 right) depending on the $R_C C$ time constant. Even though reducing the value of $R_C$ seems to be a possible solution is not always acceptable. As this would drain more power in the circuit. Here an alternate solution is proposed which improves the wave form without having to reduce the value of $R_C$. This circuit is as shown in Figure 2. This article is divided into three sections the first gives introduction to the problem, section II describes the modified circuit along with the important results of analysis of the circuit and the III[rd] section presents test results. Appendix at the end of the article documents the analysis through which results in section II were reached.

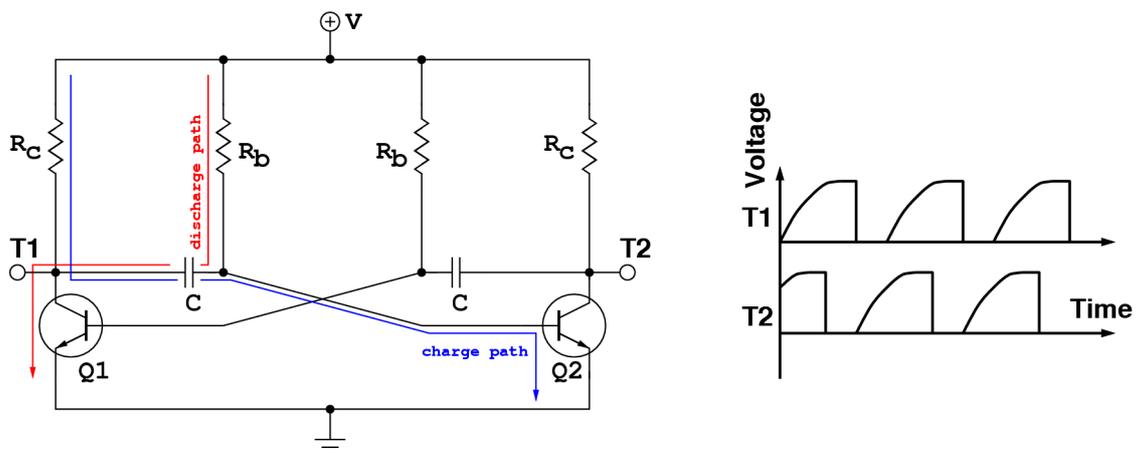

*Fig 1: Conventional multivibrator using two transistors. Voltage waveforms at the collectors of the transistors as shown to the right.*

## II. MODIFIED ASTABLE

The waveforms obtained from an ordinary transistorized astable have been improved by giving a different path for the charging of the timing capacitors. While the capacitors are discharged by the collectors through D1 and D2 respectively they will be charged through the diodes D1' and D2' respectively, avoiding the paths involving the collectors. The improved astable works exactly like the ordinary astable(Giacoletto 1977; Kasatkin & Nemtsov 1986) except for the difference in the charging path of the timing capacitors. Q3 helps to start the oscillations during power up. The final results of the analysis of the circuit have been summarised as 'on' periods of the transistors, $t_{on}^L$ and $t_{on}^S$. Superscripts: L - larger time constant & S – smaller time constant.

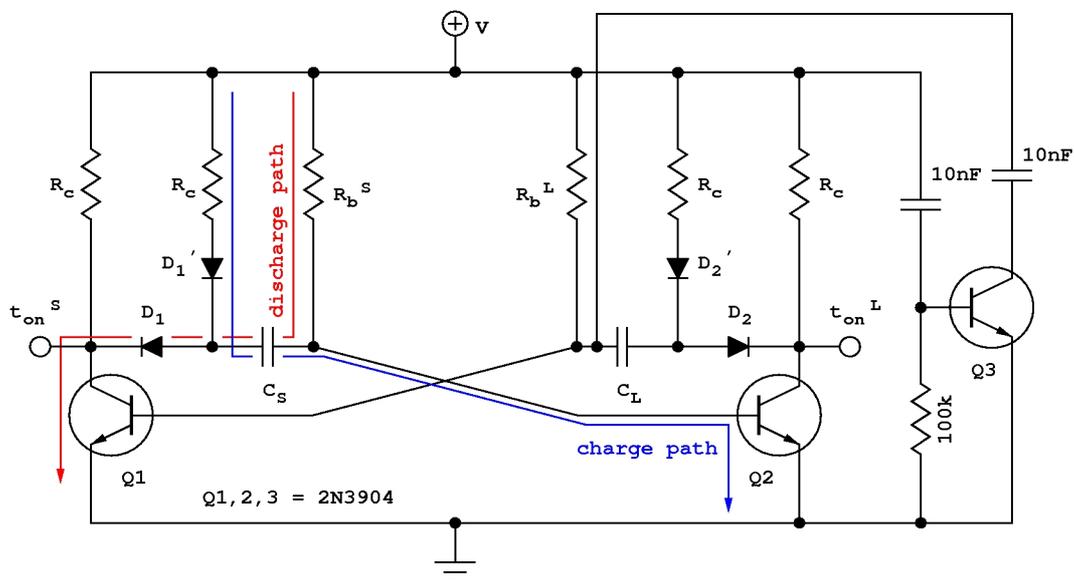

*Fig 2: Improved astable multivibrator. The diodes D1 and D2 isolate the collector from the capacitor during charging and hence the transistors turn off quickly. A separate $R_C$ branch has been provided for each capacitor for charging. Q3 helps in starting the oscillations when the circuit is powered up.*

In the symmetric case, when $R_b^S = R_b^L$ and $C_S = C_L$, $t_{on}^L = t_{on}^S = t_{on}$,

$$t_{on} \approx R_b C \ln\left(\frac{2V-(n+2)0.7}{V-0.7}\right) \approx 0.62 R_b C, \text{ at } V=5.5, n=1 \quad (1)$$

where n is the number of diodes in the $R_c$ branch, here only one (D1' and D2', D1 and D2 are for the waveform improvement alone and do not count in n).

However in the asymmetric case, due to incomplete charging of the larger time constant capacitor the above formula does not hold good due to failure of the assumption that the larger time constant capacitor was fully charged during the previous phase to the allowed value in the branch.

So in the asymmetric case,

$$t_{on}^S \approx 0.62 R_b^S C_S, \text{ for } V=5.5 \text{ and } n=1$$

$$t_{on}^L \approx R_b^L C_L \ln\left(\frac{V_{C_L}+V-0.7}{V-0.7}\right)$$

(2)

$$\text{where, } V_{C_L} = (V-(n+1)0.7)\left(1 - \left(\frac{2V-(n+2)0.7}{V-0.7}\right)^{-\frac{R_b^S C_S}{R_c C_L}}\right)$$

where the superscripts/subscripts - L and S represent the side with large and small value of time constant, $R_b C$, respectively. These formulae have been tested both by simulation and by building the circuit. They are in very good agreement with experimental values.

$$\text{frequency} = \frac{1}{t_{on}^S + t_{off}^L} \quad (3)$$

TEST VALUES: V=5.0, $R_C$=10k, $R_b$=68k, C=10nF, all diodes – 1N4148. Under simulation the circuit works even at voltages as low as 1.0 V.

## III. EXPERIMENTAL RESULTS

Simulation and test results indicate an improvement in the waveform as shown in Figure 3. The appendix discusses in detail the derivation of the formulae given in section II.

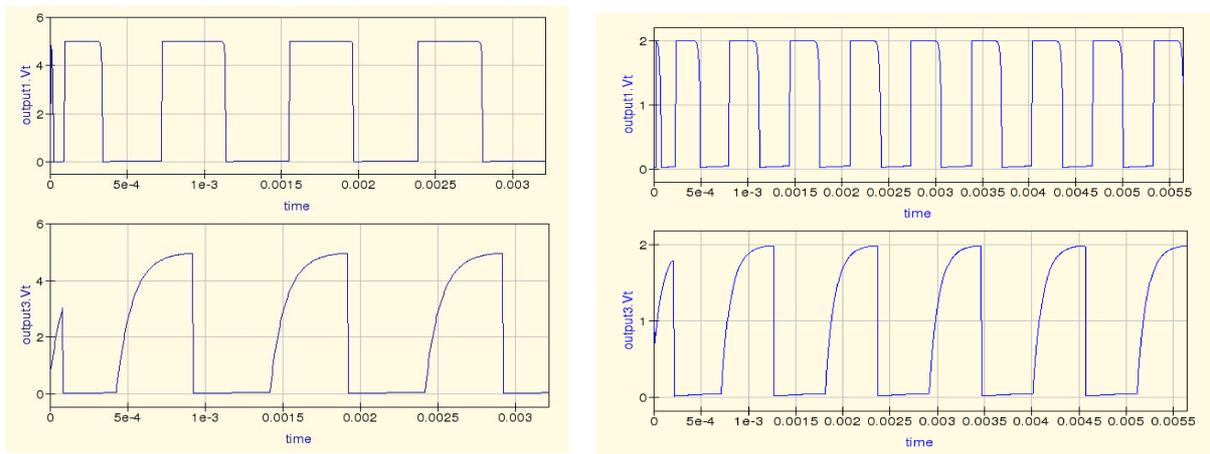

*Fig 3: Simulation plots from QUCs software in linux showing the improved waveforms. In both the panels top corresponds to the circuit improved as described here(symmetric case) where as the lower one is the traditional astable. The abscissa is time in seconds and ordinate is one of the collector outputs in volts.*

# APPENDIX

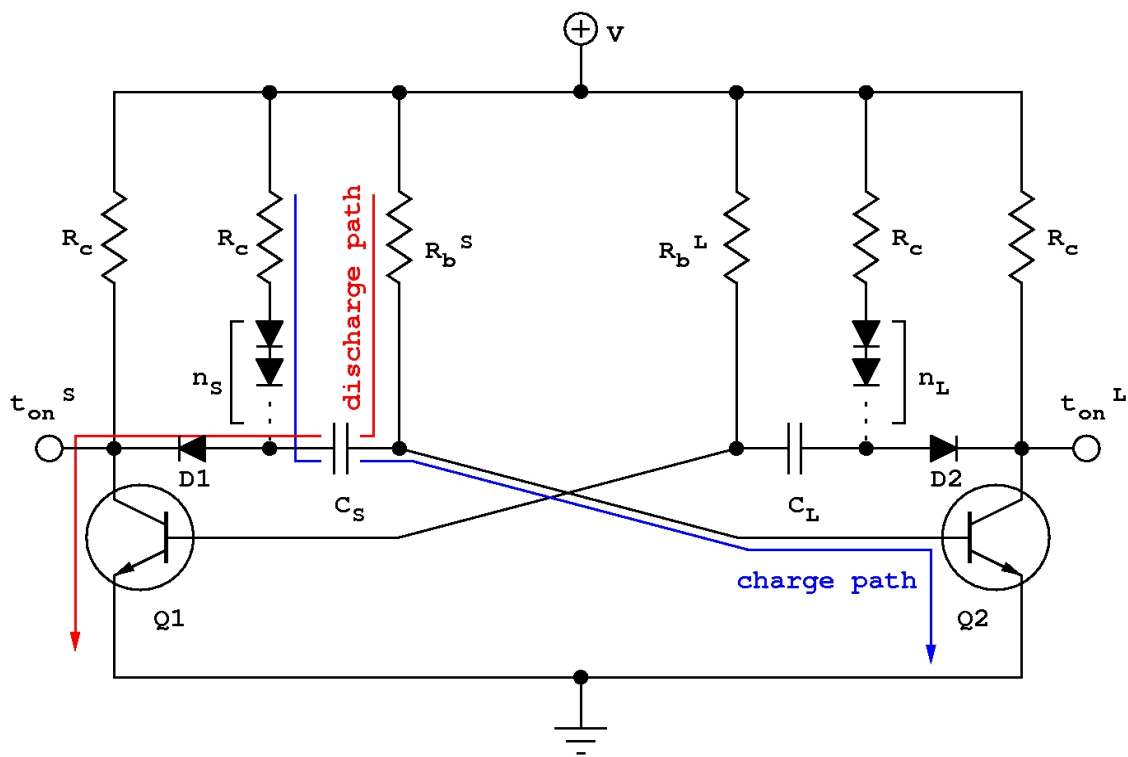

*Fig 4: Improved astable multivibrator using diodes and auxiliary collector resistances. The superscrips L and S stand for Large and Small representing larger time constant and smaller time constant sides respectively.*

We first try to understand the operation of the circuit qualitatively. One can note that the capacitors charge through $R_C$ while discharge through $R_b$. Hence the capacitor is considered to be always fully charged to the allowed value in the branch due to the low value of $R_C$ compared to $R_b$. In the schematic above the charge(blue) and discharge(red) paths have been shown. However in the case where there is an asymmetry in the values of $C_S$ and $C_L$ the charging of $C_L$ can be incomplete due to its larger value. So it follows that in the symmetric case when $C_S=C_L=C$ and $R_b^S=R_b^L=R_b$ the time period of transistors on either side can simply be taken to be the discharge time of the respective capacitors untill a swap in the on-state of the transistors occurs. As we know to turn on a transistor we need a voltage of ~0.6V across the BE junction. From the schematic it can be seen that this voltage is provided by D1 or D2 for the Q2 and Q1 respectively. For argument sake if we consider Q2 to be off and Q1 to be in the on state. Then we have a situation where a reverse charge on $C_S$ exists due to a previous state which is now decaying through the path shown in the figure. The transistor Q2 remains off untill the sum of voltages on $C_S$ and D1 are less than 0.6V. This means the capacitor $C_S$ has to fully discharge i.e $V_C^S = 0$ for Q2 to just turn on. This further simplifies the problem since we now know both the initial charge on the capacitor and the decay path for $C_S$ and like wise for $C_L$ in the symmetric case. The peak reverse charge on $C_S$ can be obtained by observing the schematic. Since $R_C$ is small we assume that the capacitor is fully charged in a short time to the residual voltage after accounting for the voltage drops across $n_S$ diodes and the BE junction of Q2.

i.e
$$V_C^S = V - (n_S+1)0.7 \tag{1}$$

Here the voltage drop across the diodes and BE junction of transistor Q2 has been chosen to be 0.7V however at lower supply voltages(< 3.0V) 0.6V may be a better choice. With this reverse charge on the capacitor $C_S$, $t_{on}^S$ is the time it takes for $C_S$ to decay to 0V through the decay path(red) shown in the figure. This decay can also be seen as charging of $C_S$ through $R_b^S$- D1- $CE_{Q1}$ starting from 0V and going to $V_C^S$. Except that the total voltage across the branch would be $(V - (n_S+1)0.7)+(V - 0.7)$. Here the CE voltage drop being small has been neglected. Hence we write for $t_{on}^S$ as,

$$V - (n_S+1)0.7 = (2V - (n_S+2)0.7)(1 - e^{\frac{-t_{on}^S}{R_b^S C_S}}) \tag{2}$$

which is rearranged to obtain $t_{on}^S$ as,

$$t_{on}^S = R_b^S C_S \ln\left[\frac{2V - (n_S+2)0.7}{V - 0.7}\right] \tag{3}$$

In the all symmetric case $t_{on}^S = t_{on}^L$. It should be noted however that asymmetry can also exist in terms of $n_S$ and $n_L$ while $C_S=C_L$ and $R_b^S=R_b^L$. In all symmetric cases $n_S=n_L=n$.

We next consider the asymmetric case where in general $C_S \neq C_L$ and $R_b^S \neq R_b^L$. We further assume that the smaller time constant side capacitor $C_S$ is fully reversed charged as in the symmetric case during the $t_{on}^L$ phase due to its smaller time constant value, $R_b^S C_S < R_b^L C_L$. We further assume that the asymmetry in $n_S$ and $n_L$ does not affect this. However in general $n_S$ and $n_L$ can also determine which side is the smaller time constant side apart from values of $R_b^S C_S$ & $R_b^L C_L$. Now we calculate the peak reverse voltage that is put on $C_L$ during $t_{on}^S$ from which we will then calcuate the decay time of $C_L$ to obtain $t_{on}^L$. As to turn on Q1 $C_L$ should have been fully discharged before Q2 turns off this simplifies the analysis and we write for the peak reverse voltage $V_C^L$ on $C_L$ as,

$$V_C^L = (V - (n_L+1)0.7)(1 - e^{\frac{-t_{on}^S}{R_C^S C_L}}) \qquad (4)$$

we can substitute (3) for $t_{on}^S$ in (4) and obtain $V_C^L$ as,

$$V_C^L = (V - (n_L+1)0.7)\left[1 - \left(\frac{V-0.7}{2V - (n+2)0.7}\right)^{\frac{R_b^S C_S}{R_C C_L}}\right] \qquad (5)$$

equation (5) gives the peak reverse voltage on $C_L$ put on it during $t_{on}^S$. We can now obtain the time period $t_{on}^L$ by the requirement that $C_L$ discharges fully through a decay loop similar to $C_S$. We write by observing the schematic a similar equation like (2) for the discharge of $C_L$ as ,

$$V_C^L = (V - 0.7 + V_C^L)(1 - e^{\frac{-t_{on}^L}{R_b^L C_L}}) \qquad (6)$$

we rearrange (6) to obtain $t_{on}^L$ as ,

$$t_{on}^L = R_b^L C_L \ln\left[\frac{V_C^L + V - 0.7}{V - 0.7}\right] \qquad (7)$$

where $V_C^L$ can be obtained from (5). We see from (4) & (7) that in the symmetric case $t_{on}^L = t_{on}^S$.

**Note :** Recommended operation $n_S = n_L$.

Next the current sink and source capabilities are discussed. When the transistor from which the signal is tapped is on the current sinks into the collector. This current should be such that the collector can still be considered to be approximately at ground potential. Since the on transistor is considered to be in a state of saturation we write for the base and collectors currents as,

$$I_b \approx \frac{V - 0.7}{R_b} \quad ; \quad I_c \approx \beta I_b \qquad (8)$$

so $I_{sink}$ is the residual available current after the requirement of $R_C$ and the connected branch are satisfied. Hence we have for this residual current as,

$$I_{sink} \approx I_C - \frac{V}{R_C} - \frac{V-(n_{S,L}+1)0.7}{R_C} \qquad (9)$$

Using (8) and rearranging we obtain ,

$$I_{sink} \approx \beta \frac{V-0.7}{R_b} - \frac{2V - (n_{S,L}+1)0.7}{R_C} \qquad (10)$$

Similarly when the transistor is off the current is sourced from the output terminal. We require this current such that the voltage drop across $R_C$ is capable of driving a current through $R_C$-D1-D1' branch. Here D1' is the set of $n_S$ or $n_L$ diodes. Since we need a limit on the voltage across $R_C$ such that a current flows through $R_C$-D1-D1' branch we choose this lower limit on the voltage to be the net voltage drop across all the diodes connecting across $R_C$. Thus we write for the source current $I_{source}$ as ,

$$I_{source} \approx \frac{(n_{S,L} + 1)0.6}{R_C}. \qquad (11)$$

**REFERENCES**

Giacoletto L.J., Electronics Designers Handbook, 2$^{nd}$ edition 1977, pp 19-13

Kasatkin A.S. and Nemtsov M.V., Electrical Engineering, Mir Publishers 1986, pp 287.